\documentstyle[12pt]{article}
\advance\hoffset -5mm 
\newcommand{\be}{\begin{equation}}
\newcommand{\ee}{\end{equation}}
\newcommand{\bea}{\begin{eqnarray}}
\newcommand{\eea}{\end{eqnarray}}
\newcommand{\vs}{{\mbox{\boldmath$S$}}}
\newcommand{\vt}{{\mbox{\boldmath$\tau$}}}
\newcommand{\vvs}{{\mbox{\boldmath$\sigma$}}}
\newcommand{\veta}{{\mbox{\boldmath$\eta$}}}
\def\lr#1{\langle#1\rangle}

\begin{document}
\baselineskip 7mm
\pagestyle{plain}
\pagenumbering{arabic}
\setcounter{page}{1}
\date{}
\title{Underscreened Kondo Necklace}
\author{P. Fazekas\thanks{Permanent address: Central Research Institute for
Physics, P.O.B. 49, Budapest 114, H--1525 Hungary}\ \ and Hae--Young
Kee\thanks{Permanent address:
Department of Physics Education, Seoul National Unversity, Seoul, 151--742
Korea}\\
International Centre for Theoretical Physics,\\
P.O. Box 586, I--34100 Trieste, Italy}

\maketitle

\vskip 1 cm

\begin{abstract}
It has been suggested recently by Gan, Coleman, and Andrei that studying the
underscreened Kondo problem may help to understand the nature of magnetism in
heavy fermion systems. Motivated by Doniach's work on the $S=1/2$ Kondo
necklace, we introduce the underscreened Kondo necklace models with $S>1/2$.
The underscreened Kondo necklace is the simplest lattice model on which the
competition between Kondo spin compensation, and magnetic ordering due to an
RKKY--type interaction can be examined. We used the mean--field approximation
to determine the phase diagram, and found that the low-temperature phase is
always an $x-y$ antiferromagnet. This contention is further supported by the
derivation of the exact form of the effective hamiltonian in the limit of
very large Kondo coupling: it is found to be an antiferromagnetic $x-y$ model
for the residual $S-1/2$ spins. In general, the degree of moment compensation
depends on both the Kondo coupling, and on $S$.
\end{abstract}

\vskip 1 cm

\section{Introduction}

The description of the magnetically ordered states of heavy fermionic systems
is a problem of great current interest \cite{GS}. It is usual to find that the
ordered
moment is rather smaller, and in quite a few cases much smaller, than what
would correspond to isolated 4$f$ (or 5$f$) shells. On the other hand, at
elevated temperatures, the paramagnetic susceptibility is the one expected for
isolated $f$--shells in a crystal field. The phenomenon of {\sl reduced moment
magnetism} at low temperatures is supposed to result from the competition of
the Kondo effect, and
RKKY interaction \cite{CG}. The Kondo effect strives to set up a state in which
the
spins of $f$--electrons are compensated by those of conduction electrons; the
RKKY interaction tends to order the localized moments, preventing the
formation of an overall singlet state. The outcome of this competition can be
apparently anything between the extremes of a conventional RKKY magnet with
well--defined localized moments, and a non-magnetic heavy Fermi liquid. We can
envisage this by saying that the
Kondo effect progresses up to a point, reducing the moments to a fraction of
the ionic value, and then the residual moments get ordered. --- The
description of this situation is greatly complicated by the fact that the
Kondo compensation clouds corresponding to nearby ions strongly overlap
\cite{{FaSh},{ShiFa}}:
there is a continuing controversy as to whether the collective Kondo effect
of the lattice has an energy scale essentially different from the single ion
Kondo energy.

The treatment of the competition between magnetic and non--magnetic states is
apparently quite difficult in the case of models with the the feature of
perfect screening: $2S=k$, where $k$ is the number of screening channels, and
$S$ is the localized spin. As demonstrated in detail \cite{FaSh}
for the case $S=1/2$,
this perfect balance allows the construction of a strictly non--magnetic heavy
fermi liquid; it remains an open question whether such a non--magnetic state
is necessarily unstable against magnetism \cite{FaMu} and/or
superconductivity \cite{CoAn}.

It has been suggested recently by Gan, Coleman, and Andrei \cite{GCA} that
one should be able to gain an insight into the nature of heavy fermion
magnetism by studying the {\sl underscreened Kondo problem} with $2S>k$.
The single--ion problem is solved exactly \cite{AD}, demonstrating that in the
ground state, a partially compensated spin $S-k/2$ remains. The underscreened
Kondo lattice (in standard notations)
\be
H_{KL} = \sum_{{\bf k}\alpha} \epsilon_{\bf k}
c_{{\bf k}\alpha}^{+}c_{{\bf }k\alpha} +
\frac{J}{L}\sum_{\alpha,\beta} \sum_{{\bf k}{\bf k}^{\prime}{\bf j}}
e^{i({\bf k}-{\bf k}^{\prime}){\bf j}}
c_{{\bf k}\alpha}^{+}\vvs_{\alpha\beta}
c_{{\bf k}^{\prime}\beta}\cdot\vs_{\bf j}
\ee
with an orbitally non--degenerate conduction band ($k=1$), and $S>1/2$, seems
destined to become magnetic (near half-filling, presumably antiferromagnetic),
but nothing specific is known yet.

Recalling that in the perfectly screened case $k=1$, $S=1/2$, the simplest
solution of the Kondo necklace model introduced by Doniach \cite{Do} has been
a useful guide to studying the more complicated problem of the Kondo lattice,
we suggest that the introduction of the {\sl underscreened Kondo necklace
models}

\be
H_{KN} = J\sum_{\bf j} \vs_{\bf j}{\cdot}\vt_{\bf j} + W\sum_{\langle{\bf i},
{\bf j}\rangle} (\tau_{\bf i}^{x}\tau_{\bf j}^{x} +
\tau_{\bf i}^{y}\tau_{\bf j}^{y})
\ee
should be helpful as a prelude to a study of the underscreened Kondo lattice.

The great simplification achieved by replacing the Kondo lattice model (1) by
the Kondo necklace model (2) is due to the fact that $H_{KL}$ has in it spins
and fermions, while $H_{KN}$ is formulated solely in terms of spin operators.
The spins $\vs$ are the same localized spins as in (1). We consider an
arbitrary spin $S$; the underscreened models belong to $S>1/2$, but to
facilitate comparison with the case of exact Kondo screening, we include also
results for $S=1/2$ which belongs to the original necklace model.

The pseudospins $\vt$, with $|\vt|=1/2$, are meant to represent the spin
degrees of freedom of the conduction electrons. Charge degrees of freedom are
omitted on the ground that they are likely to belong to higher energies, and
are thus considered to be of little relevance for the low-energy world of
Kondo physics. Note that one $\vt$--spin for each site should really correspond
to a half--filled conduction band. At this special filling, the ground state of
$H_{KL}$ is expected to be insulating: either because of Luttinger's theorem,
in case it is non--magnetic; or because of magnetic cell doubling, in case it
is an antiferromagnet. In either case, there is a charge gap, providing a
further justification for omitting the charge degrees of freedom in $H_{KN}$.

In one dimension (1$d$), the $x-y$ model of the $\vt$-spins can be mapped by
the Jordan--Wigner transformation to a tight-binding model of spinless
non-interacting fermions. In this case, it becomes quite apparent that the
$x-y$ term in (2) stands for a sea of fermions. However, even in higher
dimensions, it is clear that $H_{KN}$ incorporates the competing tendencies
of magnetic ordering, and local Kondo singlet formation. This being just the
question we are interested in, we do not hesitate to consider the model (2)
for lattices of arbitrary dimensionality and structure (even though the term
``necklace'' becomes a misnomer). --- In any case, the simple mean--field
treatment which we carry out here, in close parallel with Doniach's first
study \cite{Do} of the original necklace model, is really justified only for
3$d$, and gives at best a qualitative guide as to what to expect in lower
dimensions.

The first term in (2), with the antiferromagnetic Kondo coupling $J>0$,
favours the formation of local low-spin states with $|\vs+\vt|=S-1/2$. The
second term, with $W$ measuring, loosely speaking, the kinetic energy of the
spin degrees of freedom of the conduction electron sea, couples
nearest--neighbour sites, and thereby mixes in also components from the local
$|\vs+\vt|=S+1/2$ high--spin states.

In the perfectly screened case $S=1/2$, Doniach \cite{Do} found a ground state
phase transition between an antiferromagnetic state at $J<J_{cr}=W$, and a
fully Kondo--compensated overall singlet state at $J>W$. The antiferromagnet
has gapless spin excitations, while the Kondo state shows a spin gap. --- It
is, of course, always questionable whether mean--field results can be trusted,
and it has been a matter of long debate what the true behaviour should be
like. The problem is most subtle in 1$d$, for the true ``necklace''.
Amazingly,
the mean--field approximation turns out to be a good guide even in this case.
Naturally, true long-range order has to be replaced with quasi--long-range
order, but it seems to be true that there is a ground state phase transition
between a low--$J$ state which is almost magnetic, with the spin excitation
spectrum gapless, and a high--$J$ Kondo state where there exists a spin gap.
In finding this, very recent numerical studies \cite{SaSo} corroborate earlier
renormalization group results \cite{RNG}, refuting some Monte Carlo work
\cite{Sca} which claimed that the spectrum is gapful for all $J>0$. Note,
however, that for the 1$d$ Kondo lattice, it is reasonably expected, and also
convincingly demonstrated \cite{Tsu}, that the spectrum is always gapful.
This indicates that the breaking of the full spin--rotational invariance, which
is caused by introducing the $x-y$ form of the intersite coupling in $H_{KN}$,
puts the Kondo necklace into a different universality class than the Kondo
lattice. --- In the present context of the underscreened Kondo necklace model,
we will understand this more clearly
after deriving the effective Hamiltonian governing the large--$J$ behaviour
which turns out to be an antiferromagnetic $x-y$ model of the partially
Kondo--screened residual moments.

\section{$X-Y$ Type Antiferromagnetism in the Necklace Model}

Having thus found sufficient justification for doing simple mean--field
theory to get a first impression of the behaviour of the model (2), we
proceed to decouple the $x-y$ term, by assuming a $d$--dimensional cubic
lattice, and postulating a two--sublattice ($A$ and $B$)
antiferromagnetic order of the $\vt$-spins

\be
\lr{\tau_{i}^{x}} = \left\{
\begin{array}{ll}
t>0 & {\rm if\ } i\in A\\
-t & {\rm if\ } i\in B
\end{array}
\right.
\ee
while $\lr{\tau^{y}}=0$ everywhere.

The single--site mean--field hamiltonian is
\be
h_{MF} = J\vs\vt - 2\omega\tau^{x}
\ee
where $\omega=Wtd$, and $t=\langle\tau^{x}\rangle$ is the thermal
expectation value of $\tau^{x}$, which has to be determined self-consistently
from
\be
t = \frac{\sum_{j=1}^{4S+2} \langle\psi_{j}|\tau^{x}|\psi_{j}
\rangle e^{-\epsilon_{j}/T}}{\sum_{j=1}^{4S+2} e^{-\epsilon_{j}/T}}
\ee
Here $\epsilon_{j}$, and $|\psi_j\rangle$, are the eigenvalues, and
eigenstates of $h_{MF}$.

Conveniently chosing the quantization axis as the $x$--axis, the
$2(2S+1)$--dimensional eigenvalue
problem immediately separates into $2S$ two--dimensional, and 2
one--dimensional
problems. In the subspace $S^{x}+\tau^{x}=S-m+1/2$, where $m = 1, 2, ... 2S$,
the matrix of $h_{MF}$ is found to be
\be
\left(\begin{array}{cc}
{J\over 2}(S-m)-\omega & {J\over 2}\sqrt{m(2S-m+1)} \\[4mm]
\frac{J}{2}\sqrt{m(2S-m+1)} & -\frac{J}{2}(S-m+1)+\omega
\end{array}\right)
\ee
There is no point to writing down the solution of the diagonalization, we
proceed straight to the results obtained from solving the self--consistency
equation.

The N\'eel temperature $T_N$ is obtained from solving the linearized
self--consistency equation

\bea
\frac{Wd}{3Z_{0}(1+2S)} \Bigl \{ \frac{1}{T_N} \Bigl [
S(2S-1)e^{\frac{J(S+1)}{2T_{N}}} +
 (S+1)(2S+3)e^{-\frac{JS}{2T_{N}}} \Bigr ] + \nonumber\\
\frac{1}{J}\frac{16S(S+1)}{1+2S}\Bigl [ e^{\frac{J(S+1)}{2T_{N}}} -
e^{-\frac{JS}{2T_{N}}}\Bigr ]
\Bigr \} = 1
\eea
where
\be
Z_{0} = 2\Bigl [ Se^{\frac{J(S+1)}{2T_{N}}} +
(S+1)e^{-\frac{JS}{2T_{N}}}\Bigr ]
\ee

Some representative curves giving the antiferromagnetic--paramagnetic phase
boundary in the $J/W-T_{N}/W$ plane are shown in Fig. 1. At J=0, all
curves start from the common value $T_{N}=Wd/2$, the mean--field solution for
the $x-y$ model of the $\vt$--spins. This single point is, of course, somewhat
pathological, since the decoupled $\vs$--spins remain completely disordered
at all temperatures, including $T=0$. For all $J>0$, however, the
antiferromagnetism of the $\vt$--spins induces an (oppositely polarized)
antiferromagnetic order of the $\vs$--spins.

As the Kondo coupling reaches the range $J\approx Wd/S$, the states derived
from $|\vs+\vt|=S+1/2$ levels of the single--ion problem are beginning to
make a negligible contribution to thermal expectation values, and $T_N$
quickly drops toward its asymptotic value

\be
\lim_{{J/W}\to\infty} {T_{N}} = \frac{Wd}{6}\cdot\frac{2S-1}{2S+1}
\ee

For $S=1/2$, there is no antiferromagnetism at large $J$'s, in agreement
with the solution found by Doniach \cite{Do}. However, for all $S\ge 1$,
a low--temperature antiferromagnetic phase is predicted, $T_N$ approaching
the finite value (9), as $J/W\to\infty$. This arises from the fact the
$2S$--fold degenerate ground state set of the Kondo ion reacts with a
Curie--like polarization to the transverse field $\tau^{x}$. The argument can
be made a bit more formal by deriving the effective Hamiltonian which
expresses the action of $H_{KN}$ within the restricted Hilbert space spanned
by the $|\vs+\vt|=S-1/2$ Kondo ground states. Expressed in the basis
$|S^{z},\tau^{z}\rangle$, the relevant single--site states are

\be
\phi_{S-m-1/2} = \frac{1}{\sqrt{2S+1}} [ \sqrt{2S-m} |S-m,-\frac{1}{2}\rangle
- \sqrt{m+1} |S-m-1,\frac{1}{2}\rangle ]
\ee
where $m = 0, 1,..., 2S-1$. It is easy to check that in the Hilbert space
spanned by these local basis states, $H_{KN}$ acts like

\be
H_{eff}^{J/W\to\infty} = -J\frac{S+1}{2} + \frac{W}{(2S+1)^2}
\sum_{\langle{\bf i},{\bf j}\rangle} (\eta_{\bf i}^{x}\eta_{\bf j}^{x} +
\eta_{\bf i}^{y}\eta_{\bf j}^{y})
\ee
where the components of the usual spin operator $\veta$ for $|\veta|=S-1/2$
appear. Thus in the limit $J/W\to\infty$, the Kondo necklace model becomes
equivalent to an antiferromagnetic $x-y$ model of the residual $S-1/2$-spins,
with an effective coupling $W/(2S+1)^2$. Note that this result is rigorous,
being perfectly independent of the previous mean--field argument.

It might seem unphysical that $T_N$ approaches a finite value as $J/W\to
\infty$, since it looks obvious that it should rather tend to zero. There is,
however, no real contradiction. We have to remember that $W$ is an intersite
spin--flip matrix element of the conduction electrons, and is thus, in terms
of the underlying Kondo lattice model (1), itself dependent on $J$. In
particular, at large $J$, the $\vt$--spin--flip process involves breaking up
the local Kondo ground states, and thus $W\propto B^{2}/J$ is expected, where
$B$ is the conduction electron bandwidth. Hence (9) actually predicts
$T_{N}\to 0$ as $J/W\to\infty$.

We now return to the mean--field solution, and discuss the low--temperature
properties. Let us recall that Doniach's mean-field solution was actually
done for $T=0$, being based on a site--factorized Ansatz for the ground state
wave function. Our finite--temperature mean--field approximation should become
equivalent to this at $T=0$, since it relies on the same kind of
decoupling. This becomes quite clear from a detailed study of the $S=1$
case \cite{Kee} where both formulations were implemented. However, for
general $S$, we find it more convenient to use our present formulation.

The solution of the eigenvalue problem of (6) reveals that
the ground state is lying in the $m=2S$ subspace. The $T=0$ self--consistency
equation has no closed--form solution. However, we can carry out series
expansions, either for small, or large $J/W$. The large-$J$ expansion up to
fourth order gives
\bea
\langle\tau^{x}\rangle \approx \frac{1}{2}\frac{2S-1}{2S+1}\cdot
\{ 1 + \frac{16S}{(1+2S)^3}\frac{W}{J} -
   {16S \left( 3 - 28S + 12 {S^2} \right)
        \over \left( 1 + 2S \right)^{6}}\frac{W^2}{J^2} -
        \nonumber\\[1.5mm]
   {{128 S
       \left( 1 - 28S + 136 {S^2} - 112 {S^3} + 16 {S^4} \right)
       }\over {
       {{\left( 1 + 2S \right) }^{9}}}}\frac{W^3}{J^3} - \nonumber\\[1.5mm]
   {64S \left( 5 - 290S + 3548 {S^2} - 12976 {S^3} + 14192 {S^4} -
         4640 {S^5} + 320 {S^6} \right)  \over
      {{\left( 1 + 2S \right) }^{12}}}\frac{W^4}{J^4}\}
\eea
This is, roughly speaking, a power series in $J/WS^{2}$, i.e., in the
dimensionless effective coupling we can identify from (9).
The overall factor $2S-1$ reflects that, as known from Doniach's early
work \cite{Do}, the magnetic order is suppressed by the formation of Kondo
singlets for large $J$'s if $S=1/2$. On the other hand, no sign of any
sudden change in the nature of the ground state is detected for $S\ge 1$.

$\tau^x$ is perfectly valid as an order parameter but we rather prefer using
$S^x$, the
transverse sublattice magnetization of the $\vs$--spins. Evaluating this needs
no further calculation since, by construction

\be
\lr{\tau^{x} + S^{x}} = \pm(S-1/2)
\ee
whereby it has to be observed that ${\rm sg}(S^{x})=-{\rm sg}(\tau^{x})$.
With our convention (3), the $\vs$--spin is ``up'' on the $B$ sublattice.

The fourth--order small--$J$ expansion yields
\be
\langle\tau^{x}\rangle \approx \frac{1}{2} - \frac{S}{2}\cdot\frac{J^2}{W^2}
+ \frac{S(2S-1)}{2}\cdot\frac{J^3}{W^3} - \frac{1}{8}S(3-10S+12S^{2})
\cdot\frac{J^4}{W^4}
\ee

$S^x$ for arbitrary $J/W$ can be obtained numerically: examples are shown in
Fig. 2. Generally, $S-1/2\le S^{x}\le S$. The downward deviation from $S$ is
a measure of the strength of the (necessarily partial) Kondo compensation
which depends not only on $J/W$ but also (and rather drastically) on $S$.
For large $S$, the $\vs$--spin density wave amplitude approaches the
quasi--classical $S-1/2S$, which is accompanied by an oppositely polarized
$\vt$--spin wave of amplitude $1/2-1/2S$. This picture of a frozen state
yielded by the
mean--field treatment apparently fails to do justice to the more dynamical
internal structure of the strongly bound low--spin Kondo state, in terms of
which the effective hamiltonian (11) is formulated. --- To see where essential
features may be missed let us note that in the mean--field treatment,
``turning the spin of a site down'' is achieved by acting with a linear
combination of $S^{-}$, and $\tau^{-}$. However, the relevant operators are
the $\veta$--operators describing the composite Kondo--bound objects and in
general, $\eta^{-}$ will contain also more complicated components like
$S^{-}S^{-}\tau^{+}$, etc.

\section{Discussion and Summary}

Motivated by Doniach's \cite{Do} study of the original $S=1/2$
necklace model, we introduced the underscreened Kondo necklace model $H_{KN}$
given in (2) for studying the situation where conduction electrons in a
single screening channel can only partially compensate a lattice of localized
spins $S>1/2$. The necklace model incorporates only one aspect of the full
Kondo lattice problem: the competition between Kondo compensation, and
intersite spin exchange, and is thus meant to model magnetic ordering in the
case when the (non--degenerate) conduction band is half--filled, and the
system is expected to have an insulating ground state. Though the denomination
``necklace'' is suggestive of 1$d$, the above--described situation is present
in all dimensions, and we have in mind really three--dimensional systems.
Within a mean--field
approximation, we found that the low--temperature phase is a two--sublattice
$x-y$ antiferromagnet for arbitrary strength of the Kondo coupling $J$. It
also follows that the excitation spectrum is gapless, in the same sense as the
ordered state of the original necklace model was found to be.

Obviously, the Kondo necklace models do not have real ``Kondo physics'' in
them, in the sense that low--energy electron--hole excitations, and with them
the possibility of the characteristic non--analyticities at weak coupling, are
left out. Kondo divergencies are expected to be most important when the
(in the lattice case, collective) Kondo effect hinders, or at least greatly
suppresses, ordering. Such is not the case with the underscreened necklace
models where we find well--developed antiferromagnetism. It can be argued
\cite{LiC} that in the cases of clearcut ordering, the eventual inclusion of
true Kondo physics would shift the phase boundaries, but would not modify the
overall appearance of the mean--field phase diagram. Our results should be
considered in this spirit.

In addition to missing out on the charge degrees of freedom, the necklace
models differ in another significant way from the Kondo lattice models:
they turn out to describe $x-y$ type magnets rather than isotropic ones.
This was already implicit in the fact that Doniach's \cite{Do} mean field
trial state turned out to be polarized along the spin--$x$ axis, but some
subsequent studies \cite{Sca} seemed to hint that the Kondo necklace may be,
after all, in the same universality class as the isotropic Kondo chain.
This question is bound up with that of the existence of a spin gap in the
spectrum. There seems to be no doubt that the spin excitation spectrum of the
$1d$ Kondo lattice is gapful \cite{Tsu}. The recent numerical finding
\cite{SaSo} of a phase transition between a gapless and a gapful state in the
ground state of the $S=1/2$ Kondo necklace underlines two points: a) that the
anisotropic form of the $\vt$--term in (2) has essential consequences, and
b) that the mean field treatment is a useful guide to finding out whether
there is such a phase transition.

In the underscreened necklace models, the $x-y$ character becomes even more
emphatic. We derived the exact form (11) of the effective hamiltonian governing
the large--$J$ behaviour, and found it to be a pure antiferromagnetic $x-y$
coupling between the residual composite spins. --- From this it also follows
that -- in contrast to the expected behaviour of the $1d$ underscreened Kondo
lattice models -- the Kondo necklaces show no Haldane phenomenon, i.e., the
nature of the ground state is not expected to alternate between integer and
half--integer values of $S$. This should be clearest at $J/W\to\infty$ when
(11) can be used. In the ground state phase diagram of a larger class of spin
models \cite{KeTa}, the pure antiferromagnetic $x-y$ model is lying on the
boundary between the Haldane, and $x-y$ phases, and is supposed to do nothing
exotic.

The results for the ground state sublattice magnetization are given in Fig.~2.
It is a question of great current interest \cite{{GCA},{CG}}, how the
$f$-spins get apparently divided into a screened, and an ordered part. The
simplification
brought by replacing the Kondo lattice model with the necklace model allows a
simple treatment  of this problem. As $J/W$ is increased, the order parameter
shows a perfectly smooth behaviour, gradually decreasing from its unscreened
value at $J=0$, towards the asymptotic value (9).

To summarize, to gain some insight into the difficult problem posed by the
underscreened Kondo lattices \cite{GCA}, we introduced the underscreened
Kondo necklace models (2). These are in the same relationship to the Kondo
lattices as Doniach's \cite{Do} original necklace model to the $S=1/2$ Kondo
lattice. The simplification has its price: charge degrees of freedom are not
considered, and the spin-rotational invariance of the original Kondo lattice
model is destroyed. By deriving the exact form (11) of the effective
hamiltonian governing the behaviour of the lattice model at very large Kondo
coupling, we found that it describes an $x-y$ antiferromagnet of the residual
spins. In an effective field treatment, the ground state is always
antiferromagnetic, and the spectrum gapless. The mean field phase diagram is
shown in Fig. 1. Though obviously oversimplified, the model introduced by us
has the merit of allowing the description of the competition between Kondo
compensation and magnetic ordering for underscreened localized moments.

{\bf\large Acknowledgements} The authors wish to express their gratitude
to the International Centre for Theoretical Physics for financial support,
hospitality, and an encouraging scientific atmosphere. P.F. is grateful to
J. S\'olyom for useful advice concerning one
point in the argument, and J. Gan for sending a copy of his Thesis.


%
%
\newpage
{\bf\large Figure Captions}
\vskip .8 cm

{\sl Fig. 1} N\'eel temperature $T_N$ in units of $W$ versus $J/(J+W)$, with
$d=1$, for several values of $S$. Note that for the underscreened models
$(S>1/2)$, $T_{N}/W$ tends to a finite value as $J/W\to\infty$.
\vskip .4cm
{\sl Fig. 2} The $T=0$ value of the order parameter $S^x$ changes in the
range $[S-1/2,S]$ as a function of $J/W$ ($d=1$ was taken). For the
underscreened $S>1/2$ models, no ground state phase transition is found.
\end{document}